# New apparatus for DTA at 2000 bar: thermodynamic studies on Au, Ag, Al and HTSC oxides

V.Garnier[1,3], E. Giannini[1], S. Hugi[2], B. Seeber[2], R. Flükiger[1]

1 DPMC, Université de Genève, 24 quai E.-Ansermet, 1211 Genève 4, Switzerland
2 GAP, Université de Genève, 20 rue de l'école de médecine, 1211 Gèneve 4, Switzerland
3 Present address: INSA-GEMPPM Bat. Blaise Pascal, 20 avenue Albert Einstein, 69621 Villeurbanne, France.
Vincent.Garnier@insa-lyon.fr

**Abstract**

A new DTA (Differential Thermal Analysis) device was designed and installed in a Hot Isostatic Pressure (HIP) furnace in order to perform high-pressure thermodynamic investigations up to 2 kbar and 1200 °C. Thermal analysis can be carried out in inert or oxidising atmosphere up to $p(O_2) = 400$ bar. The calibration of the DTA apparatus under pressure was successfully performed using the melting temperature ($T_m$) of pure metals (Au, Ag and Al) as standard calibration references. The thermal properties of these metals have been studied under pressure. The values of $\Delta V$ (volume variation between liquid and solid at $T_m$), $\rho_{sm}$ (density of the solid at $T_m$) and $\alpha_m$ (linear thermal expansion coefficient at $T_m$) have been extracted. A very good agreement was found with the existing literature and new data were added.
This HP-DTA apparatus is very useful for studying the thermodynamics of those systems where one or more volatile elements are present, such as high $T_C$ superconducting oxides. DTA measurements have been performed on Bi,Pb(2223) tapes up to 2 kbar under reduced oxygen partial pressure ($p(O_2) = 0.07$ bar). The reaction leading to the formation of the 2223 phase was found to occur at higher temperatures when applying pressure: the reaction DTA peak shifted by 49 °C at 2 kbar compared to the reaction at 1 bar. This temperature shift is due to the higher stability of the Pb-rich precursor phases under pressure, as the high isostatic pressure prevents Pb from evaporating.

## I – Introduction

Material science requires a deep knowledge of the phase diagrams of elements and compounds as well as the understanding of phase transformations and the control of reaction kinetics. Thermodynamic studies provide the basic tool for material processing. Thermal analysis, which was developed more than one century ago and is widely used to study the thermodynamic properties of elements and compounds, is nowadays a fast and powerful technique to study phase diagrams and phase transformations in novel materials and complex systems like oxide superconductors or organic compounds. Phase transitions or reactions induce temperature changes in the sample due to negative (exothermic reaction) or positive (endothermic reaction) enthalpies. In thermal analysis, the temperature changes in the sample are recorded as a function of time, either upon heating or cooling. This experimental technique was performed first by Le Chatelier in 1887 [1]. However, the small temperature deviations which occur in a sample cannot be detected using this technique. In 1889, Roberts-Austen [2] proposed that differential measurements be performed. By using two thermocouples, and placing one in contact with the sample and the other in contact with an inert reference in the same furnace, a temperature difference between them could be read as soon as a reaction occured in the sample. Thus, the differential temperature reading, which is more sensitive to small temperature changes in the sample than the single thermocouple method, is recorded as a function of time or temperature. Nowadays, DTA is a widespread and well-known method for thermodynamic investigations.

However, thermodynamic data are more difficult to obtain under high pressure. Measurements of the melting points of various substances under pressure have been performed by Tammann [3] and Bridgman [4, 5] by recording the discontinuity in the volume at the transition. The detection and measurement of the melting point of a metal under pressure has also been determined from the latent heat step in a rising time-temperature curve [6] (thermal analysis). More recently, the DTA technique has been employed using various gas atmospheres at different pressures [7 - 9]. Using a pressurised gas is better than inducing pressure with anvils. Pure metals can often be studied by both methods, but multi-component systems are in general sensitive to the composition of the atmosphere (mainly oxygen content) and thus investigations of these systems require gas pressures to remain as close as possible to the synthesis conditions.

In this article, we report the development and the installation of a DTA head inside a HIP (Hot Isostatic Pressure) furnace. The calibration of our DTA apparatus was performed using Au, Ag and Al.. This work allows us to measure the fundamental thermal parameters of gold, silver and aluminium under pressure. New data, not available in the literature so far, have been obtained on these metals. The effect of pressure on Bi,Pb(2223) phase formation has been investigated by means of our new set-up, by measuring Bi,Pb(2223)/Ag tapes.

## II – Experimental

The DTA head used in this work was designed by modifying a commercial three-sensor DTA sample-holder. The DTA head is held inside a double alumina screen in order to avoid thermal gradients and fluctuations due to the gas convection (Figure 1). The thermocouples coming out from the DTA head are welded above the alumina screen in order to extend the connection out of the furnace. Care was taken to avoid any spurious thermoelectric power on the differential thermal signal. This DTA head was then installed in a HIP furnace consisting of a cylindrical high-pressure chamber (height 405 mm, diameter 154 mm, volume 7.5 litres) able to operate under oxidising atmosphere up to 2 kbar and 1200 °C. The chamber is equipped with four thermocouples, two for the furnace heating control (furnace thermocouples) while the other two are in contact with the DTA head in the middle of the furnace (sensor thermocouples) (Figure 2).

Pressure gradients affects the temperature measurement which is of prime importance in a DTA [10]. The effect of pressure on the output of several technological thermocouples has been measured by many groups (see [11] for a summary). Results strongly depend on the kind of thermocouple. The most accurately characterised thermocouples at high temperature are chromel-alumel (C/A, type K) and platinum-platinum/10% rhodium (P/P10Rh, type S). Temperature corrections ($\Delta T$) under pressure for C/A and P/P10Rh are of opposite sign. $\Delta T$ must be added to the measured temperature using calibration curves at 1 bar for P/P10Rh, but subtracted for C/A. However, these corrections are small ($\Delta T/\Delta P < 0.5$ °C·kbar$^{-1}$ for T<1000 °C) [12]. A P/P10Rh thermocouple is used in our DTA device, and using the data of Lazarus *et al*. [12], a correction of $\Delta T/\Delta P \sim 0.45$ °C·kbar$^{-1}$ for 750<T<1000 °C was applied to our measurements. As our DTA apparatus can reach a maximum of 2000 bar, the largest correction which was made was $\Delta T = 0.9$ °C (P = 2000 bar, T = 1000 °C).

For all the DTA experiments presented in this paper, a heating rate of 5 °C/min was chosen. High purity samples were used for the calibration study: gold (Goodfellow 99.95%), silver (Goodfellow 99.95%) and aluminium (Goodfellow 99.999%). Six different pressures have been tested for each element, ranging from 1 to 2000 bar. The gas used as a pressure transmitter is 99.998% pure argon. For our DTA experiments on $(Bi,Pb)_2Sr_2Ca_2Cu_3O_{10+\delta}$ (Bi,Pb(2223)), we performed the HIP-DTA measurement on multifilamentary unreacted tapes cut into short lengths in order to completely fill the DTA crucible. The oxygen partial pressure $p(O_2)$ was fixed at 0.07 bar whatever the total applied pressure, i.e. starting with a mixture of 700 ppm $O_2$ in argon to get 0.07 bar $O_2$ at 100 bar total pressure. The oxygen partial pressure in the HIP furnace before and after treatment was found to be unchanged.

## III – Results and discussion

### III – 1 - Theoretical aspect of dP/dT calculation

The dependence of the melting point of a given element or compound on the pressure is described by the Clausius Clapeyron equation,

$$\frac{dP}{dT_m} = \frac{\Delta H_m}{T_m \Delta V_m} \qquad (Eq1)$$

where $\Delta H_m$ is the melting enthalpy, $\Delta V_m$ is the volume variation between liquid and solid at the melting temperature $T_m$. $\Delta H_m$ values are well known for pure elements, and reference values for silver, gold and aluminium are reported in table 1. On the other hand $\Delta V_m$ is much more difficult to find and large variations between data are reported in the literature *(see section III-3)*. The $\Delta V_m$ variation is calculated as follows,

$$\Delta V_m = V_{lm} - V_{sm}$$

where $V_{lm}$ and $V_{sm}$ are the volume of the liquid and solid phase at the melting point, respectively. $V_{lm}$ can easily be calculated using the value of $\rho_{lm}$ (density of the liquid at the melting point) as given in the Handbook [13] (see table 1), but the volume of the solid at the melting point, $V_{sm}$ has to be found using a different method. $V_{sm}$ can be calculated using the following equation,

$$dV = V_{sm} - V_{298K} = V\beta dT - V\kappa dP \qquad (Eq2)$$

where $dV$ is the difference between the volume of the solid at the melting point ($V_{sm}$) and the volume at RT, $\beta$ the thermal expansion coefficient and $\kappa$ the compression coefficient of the material. Although heating and compression occur simultaneously, the volume is a state function and one can break the process into two sequential steps, one isothermal and one isobaric, in order to easily calculate the volume change.

For the isothermal step: $dT=0$: $\qquad V_{dT=0} = V_{298K}(1 - \kappa(P_{final} - P_0))$

For the isobaric step: $dP=0$: $\qquad V_{sm} = V_{dT=0}(1 + \beta(T_m - T_0))$

where $P_0$ and $T_0$ are the standard RT values. The coefficient of cubic expansion is $\beta \approx 3\alpha$, where $\alpha$ is the coefficient of linear thermal expansion for a cubic crystalline form (eg.: Au, Ag and Al). Considering that the volume compression is very small even at the highest pressures used in this study (for aluminium the volume compression is less than 0.3% at 2000 bar [14]), the volume changes during the isothermal compression were assumed to be negligible.

Even if a slight increase of the melting enthalpy ($\Delta H_m$) is observed under very high-pressure [15], $\Delta H_m$ was assumed to be constant at the pressures used in our experiments.
Because of the discrepancies among the sources available in the literature *(see section III-3)* concerning the values of β and κ, quite a large range of calculated $\Delta V_m$ can be obtained and the reliability of these calculations is uncertain. Therefore, the comparison of our experimental results with the literature was found to be difficult.

## III – 2 - Experimental results

Figure 3 shows DTA curves of gold obtained under pressures up to 2000 bar. The melting temperatures, at different pressures, for Au (deduced from the measured DTA curves of the figure 3) and for Ag and Al are summarised in table 2. The error on the measured melting temperatures is estimated using a least square method applied on both the base line and the linear fit at the beginning of the exothermic peak. The dependence of the melting temperature of Au, Ag and Al on pressure is reported in Figure 4. The melting temperature increases with pressure as predicted by the Clausius-Clapeyron law and follows a linear trend up to 2000 bar. However, before applying equation 1 to the experimental data, the temperature correction due to the pressure effect on thermocouples was taken into account. As discussed in section II, a negative correction $\Delta T/\Delta P \sim 0.45$ °C·kbar$^{-1}$ [12] was applied. Then, the dP/dT Clausius-Clapeyron slope was calculated using the least square method to give 178 bar·K$^{-1}$ for Au, 165.4 bar·K$^{-1}$ for Ag and 166 bar·K$^{-1}$ for Al (Table 2). The corrected melting temperatures obtained after applying Eq. 1 are shown in figure 5 and deviate by less than one degree from the theoretical values.

## III – 3 - dP/dT calculation and discussion

In order to check that experimental results are correct and thus to prove that our DTA calibration is suitable to study new thermodynamic systems, we need to compare the experimental slope with dP/dT calculated from $T_m$, $\Delta H_m$ and $\Delta V_m$ values found in the literature. Although $T_m$ and $\Delta H_m$ are well known for pure metals, $\Delta V_m$ is difficult to find as discussed above. As an example, for Au: $\Delta V_m/V$ can vary from 3.4% [16] to 7.1% [17]. One way to avoid these discrepancies is to calculate $\Delta V_m$ with $\rho_{lm}$ and $\rho_{sm}$, the density of the liquid at $T_m$ and the density of the solid at $T_m$, respectively. Values of $\rho_{lm}$ for pure elements are known, but only a few authors have reported values of $\rho_{sm}$ and discrepancies exist. For example, for Ag, $\rho_{sm}$ was found to vary from 9.73 [18] to 9.85 g/cm$^{-3}$ [19]. Nevertheless, this density can be calculated as follows (Eq3)

$$\rho_{sm} = \frac{\rho_{298K}}{1 + 3\alpha_m(T_m - 298)} \tag{Eq3}$$

where $\rho_{sm}$ is the density of the solid at the melting temperature $T_m$, $\rho_{298K}$ is the density at 298K (see table 1) and $\alpha_m$ is the linear thermal expansion coefficient at $T_m$. Unfortunately, values of α are available at lower temperatures only (usually up to 700 K) and some disagreement among them is found in the literature. By extrapolating to $T_m$ quite a large error on α is obtained, as an example, for Ag, α could vary from 26.8 10$^{-6}$K$^{-1}$ [20] to 32.5 10$^{-6}$K$^{-1}$ [21].

Calculated values of dP/dT$_m$ for gold, silver and aluminium obviously show strong variations depending on the choice of the above mentioned data. All these discrepancies between the data confirm that no reliable data are available to know the melting point dependence with pressure.

First, we decided to compare our experimental $dP/dT_m$ values with the scarce sources available in the literature. For Al: our value is 6.2% higher than the one measured by Butuzov [22] (156.2 bar·K$^{-1}$) and 4.6% higher than the one measured by Gonikberg [23] (158.7 bar·K$^{-1}$). For Ag, our value is 9% lower than the one measured by Kennedy and Newton [9] (181.8 bar·K$^{-1}$). For Au, Chino [24] evaluates the increase of gold's melting point as 16.6 °C under 2940 atm, which corresponds to 177 bar·K$^{-1}$; our value is less than 1% above this. Considering that our experimental results are of the same order of magnitude as those obtained in the literature, the calculation of the specific values such as $\Delta V$, $\rho_{sm}$ and $\alpha_m$ was performed using our experimental results, and the results are shown in Table 3.

Even if $\Delta V$, $\rho_{sm}$ and $\alpha_m$ are not obtained independently (see Eq. 2 and 3), it is interesting to compare each value separately with those found in the literature, which are usually given without the other ones necessary to calculate either $\Delta V$, $\rho_{sm}$ or $\alpha_m$.

To the best of our knowledge and taking into account a recent paper on the measurement of the bulk density of metals in the solid and molten regions [25], very little reliable data exists for the density of metals during melting and in the liquid regions. For the elements we studied no experimental data have been found, but two calculated densities of the solid at the melting point have been determined for Ag and Al by Pascal *et al*. [26] and Kirk-Othmer [27], and are found to be 9.85 and 2.55 g·cm$^{-3}$, respectively. Our measurements give 9.79 g·cm$^{-3}$ for Ag and 2.53 g·cm$^{-3}$ for Al, which is in very good agreement (< 1%).

*Au*: Concerning $\Delta V/V$, only old values are available in the literature for Au. At the beginning of this section, the large variation between reported values had already been exposed for gold: from 3.4% [16] to 7.1% [17], and our calculated value of 4.86% is an average of these two values. Furthermore, other $\Delta V/V$ data, 5.08%, 5.17% and 5.19% which were measured by [28], [29] and [18] respectively, show small deviations from 4.5% to 6.4% compared to our calculated value.

*Al*: We report three $\Delta V/V$ values from the literature for Al: 6.26% [29], 6.41% [30] and 6.5% [27]. Our calculated $\Delta V/V$ value, 6.48%, is very close to the most recent one found in the literature and confirms the reliability of our measurement.

*Ag*: A $\Delta V/V$ value of 5.02% have been deduced from our measurement for Ag. This value is essentially identical to the two others found in the literature: 4.99% [18] and 5% [31].

The linear thermal expansion coefficient, $\alpha_m$, at $T_m$ is more commonly found in the literature, but further discrepancies exist concerning these values for Au, Al and Ag. Our measured $\alpha_m$ values are 20.3 10$^{-6}$K$^{-1}$, 35.5 10$^{-6}$K$^{-1}$ and 25.9 10$^{-6}$K$^{-1}$ for gold, aluminium and silver, respectively. Depending on the chosen reference, our measured values could be either higher than others (18.5 10$^{-6}$K$^{-1}$ (gold) [20], 30 10$^{-6}$K$^{-1}$ (aluminium) [32]) or lower (23.07 10$^{-6}$K$^{-1}$ (gold) [33], 39.6 10$^{-6}$K$^{-1}$ (aluminium) [33], 27.05 10$^{-6}$K$^{-1}$ (silver) [20]). It should be noted that the values of $\alpha$ reported in the literature are usually obtained by extrapolating lower temperature data (<700 °C) up to the melting temperature and large deviations are expected. In spite of these uncertainties, the literature values are close to our measured values from -4 % to 15 %. We observed that our values are, in general, lower than those found by extrapolation in other papers. But, if we consider the coefficient of cubic expansion ($\beta_m \approx 3\alpha_m$), which has been measured at $T_m$ by Kubaschewski [34] without any extrapolation, a better agreement is found with our values. For Au, our value is 60.9 10$^{-6}$K$^{-1}$ compared with the calculated one of 58 10$^{-6}$K$^{-1}$, for Ag we obtained 77.7 10$^{-6}$K$^{-1}$ compared with 81 10$^{-6}$K$^{-1}$ and for Al our result is $106.5 \ 10^{-6}K^{-1}$ compared to $99 \ 10^{-6}K^{-1}$, giving a relative difference of 4 to 7% between our results and the ones of Kubaschewski [34].

Even if large data scattering is found in the literature, a good agreement has been established with our experimental results, showing that our DTA cell is well calibrated and suitable to study new systems accurately.

### III – 4 – Application to the $(Bi,Pb)_2Sr_2Ca_2Cu_3O_{10+\delta}$ system (Ag sheathed tape)

The first high-pressure investigations of the $(Bi,Pb)_2Sr_2Ca_2Cu_3O_{10+\delta}$ phase diagram were recently performed by Lomello-Tafin *et al* [35]. Information concerning the Bi,Pb(2223) phase stability was obtained up to 100 bar, showing the importance of high pressure to achieve a thermodynamic equilibrium in forming the Bi,Pb(2223) phase. The high isostatic pressure was found to prevent Pb from evaporating, and kept the actual stoichiometry closer to the nominal one. Furthermore, high isostatic pressure is successfully used to increase the core density and the critical current density of Bi,Pb(2223) tapes [36, 37]. For these reasons there is a strong interest in studying the thermodynamics of the Bi,Pb(2223) system under pressure.

Figure 6 shows DTA curves of a Bi,Pb(2223) green tape acquired under three different isostatic pressures ($P_{tot}$=1, 100, 2000 bar) at $p(O_2)$=0.07 bar. The peak's area decreases when the total pressure increases from 100 bar to 2000 bar; this behaviour could be explained by the decrease of the liquid phase content while increasing the pressure. The endothermic peaks are due to the decomposition of the $Ca_2PbO_4$ and the Bi,Pb(2212) phases, and as the stability of these lead containing phases changes under pressure, the shape of the peak changes correspondingly. Therefore, the increase of the partial melting temperature due to the increase of the pressure, from 830 °C at 1 bar to 879 °C at 2000 bar, is interpreted in terms of higher lead containing phases stability when pressure increases; the high isostatic pressure prevents Pb from evaporating. The mass loss was found to be reduced under high pressure from 1.14% to 0.31% for 100 and 2000 bar respectively, resulting in a higher lead content in the Bi,Pb(2223) tape, closer to the nominal stoichiometry.

The study of the $(Bi,Pb)_2Sr_2Ca_2Cu_3O_{10+\delta}$ system under pressure is still in progress, and complete results will be published elsewhere in the near future.

## IV – Conclusion

A new high pressure DTA has been designed and its calibration has been successfully performed using gold, silver and aluminium. The dependence of the melting point on pressure for the pure metals we chose for calibration has been verified using the Clausius Clapeyron equation. Furthermore, specific thermal parameters of the elements, such as $\Delta V$, $\rho_{sm}$ and $\alpha_m$ have been calculated and compared to the literature values and a good agreement was found The first DTA measurements up to 2000 bar on Bi,Pb(2223) tapes have been performed the oxygen partial pressure $p(O_2)$ being fixed at 0.07 bar. An increase of 49 °C of the Bi,Pb(2223) melting temperature in Ag-sheathed tapes was observed at 2000 bar compared to the value at 1 bar.

**Figure captions**

**Table 1:** Theoretical values for Ag, Al and Au [13].

**Table 2:** Melting temperatures with errors at different pressures for Au, Ag and Al (deduced from the experimental DTA curves).

**Table 3:** Calculation of specific thermodynamic values for Au, Ag and Al using our experimental results.

**Figure 1:** Picture of the DTA head.

**Figure 2:** Vertical section of the HIP furnace.

**Figure 3:** DTA curves of Au at various pressures.

**Figure 4:** Dependence of the melting temperature of Au, Ag and Al on pressure.

**Figure 5:** Corrected melting temperatures of Au, Ag and Al.

**Figure 6:** DTA curves of a Bi,Pb(2223) green tape at various pressures. The start of the partial melting temperatures 830 °C, 838.5 °C and 879 °C are indicated by arrows for 1, 100 and 2000 bar respectively.

|  | **Silver** | **Aluminium** | **Gold** |
|---|---|---|---|
| $H_{melting}$ [kJ·mol$^{-1}$] | 11.3 | 10.71 | 12.55 |
| $T_{melting}$ (at 1 bar) [°C] | 961.78 | 660.32 | 1064.18 |
| $\rho_{(25 °C)}$ [g·cm$^{-3}$] | 10.5 | 2.7 | 19.3 |
| $\rho_{(liq. melting)}$ [g·cm$^{-3}$] | 9.32 | 2.375 | 17.31 |
| $M_{Ag}$ [g·mol$^{-1}$] | 107.87 | 26.98 | 196.97 |

TABLE1

| **Gold** 178.12 bar·K$^{-1}$ 5.614 K·kbar$^{-1}$ | Pressure (bar) | 1 | 101 | 509 | 1010 | 1507 | 2012 |
|---|---|---|---|---|---|---|---|
| | Measured temperature (°C) | 1064.3 | 1065.5 | 1068.3 | 1070.9 | 1073.9 | 1076.9 |
| | Error (°C) | 0.8 | 0.8 | 0.7 | 0.7 | 0.8 | 0.7 |
| **Silver** 165.38 bar·K$^{-1}$ 6.047 K·kbar$^{-1}$ | Pressure (bar) | 7 | 102 | 509 | 1004 | 1513 | 2006 |
| | Measured temperature (°C) | 961 | 962 | 964.9 | 968 | 971.4 | 974 |
| | Error (°C) | 1 | 1 | 0.9 | 1 | 0.9 | 1 |
| **Aluminium** 165.95 bar·K$^{-1}$ 6.026 K·kbar$^{-1}$ | Pressure (bar) | 1 | 103 | 512 | 1003 | 1515 | 2008 |
| | Measured temperature (°C) | 659.6 | 660.2 | 662 | 665.8 | 669.4 | 672.5 |
| | Error (°C) | 0.5 | 0.5 | 1 | 0.5 | 0.5 | 0.5 |

TABLE2

|  | **Gold** | **Silver** | **Aluminium** |
|---|---|---|---|
| **dP/dT** [bar·K$^{-1}$] | 178.12 | 165.4 | 165.95 |
| **ΔV** [cm$^3$·mol$^{-1}$] | 0.527 | 0.553 | 0.691 |
| **ρ$_{sm}$** [g·cm$^{-3}$] | 18.15 | 9.788 | 2.53 |
| **ΔV / V$_{sm}$** [%] | 4.86 | 5.02 | 6.48 |
| **α$_m$** [10$^{-6}$·grd$^{-1}$] | 20.3 | 25.9 | 35.5 |

TABLE3

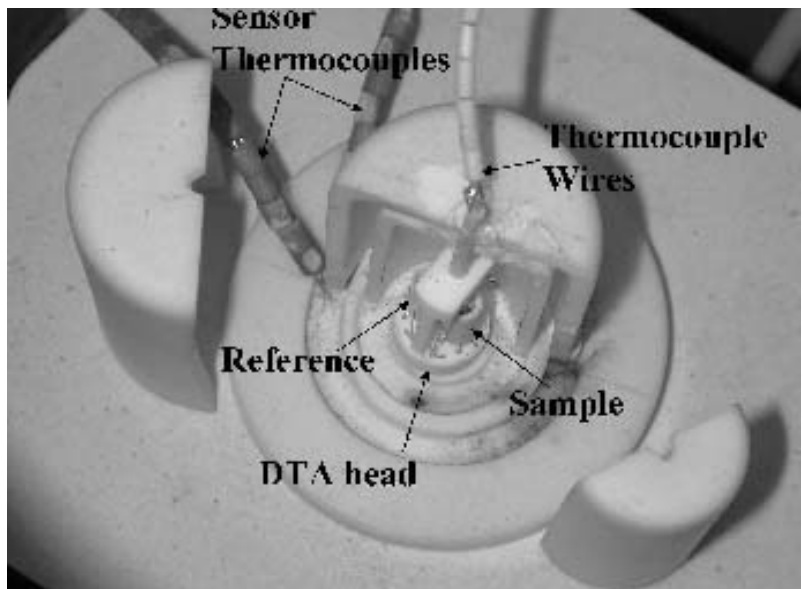

FIGURE1

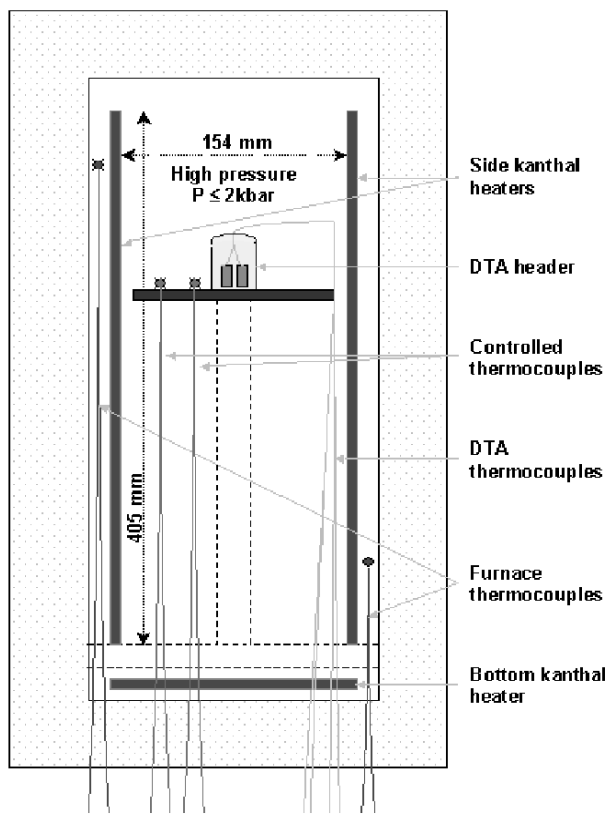

FIGURE2

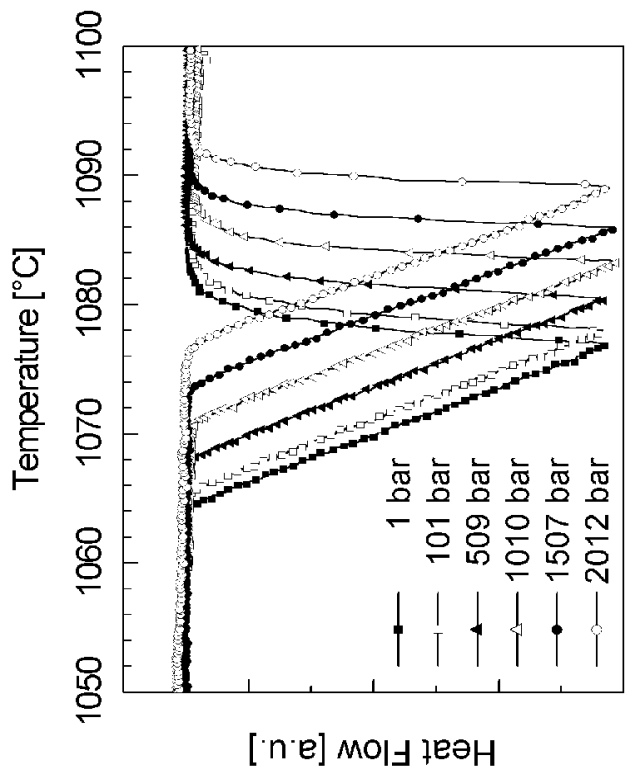

FIGURE 3

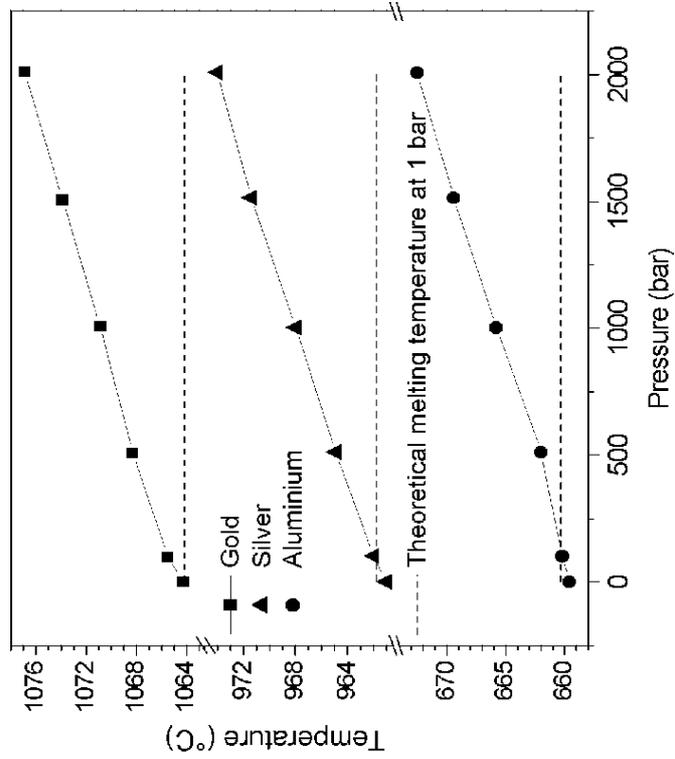

FIGURE 4

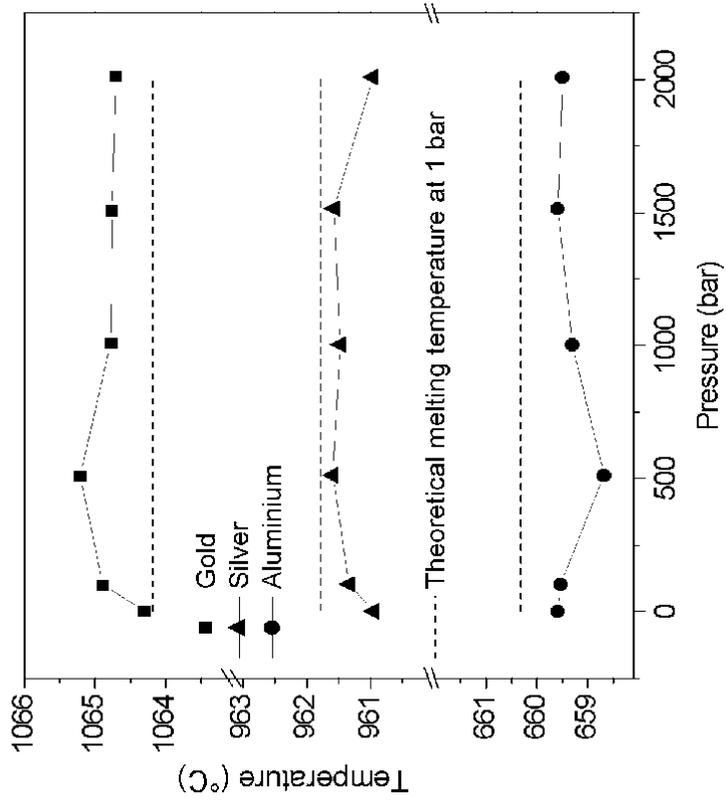

FIGURE 5

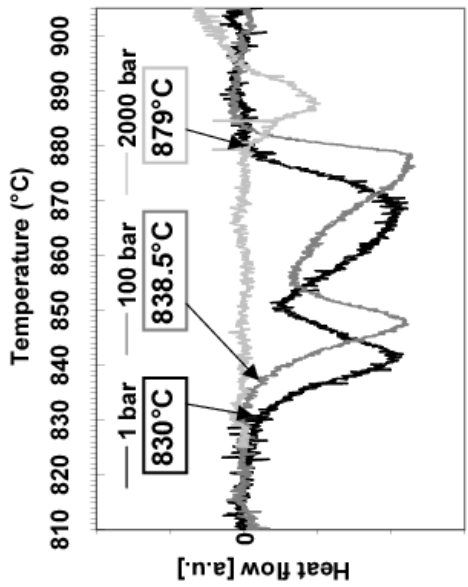

FIGURE 6